\documentclass[12pt]{iopart}
\usepackage{graphicx} 
\begin{document}

\title[Effect of base-pair sequence on B-DNA thermal conductivity]{Effect of base-pair sequence on B-DNA thermal conductivity}

\author{Vignesh Mahalingam$^1$ and Dineshkumar Harursampath$^1$}

\address{$^1$Department of Aerospace Engineering, Indian Institute of Science, Bengaluru-560012, India}
\ead{vigneshm@iisc.ac.in and dineshkumar@iisc.ac.in}
\vspace{10pt}
\begin{indented}
\item[]April 2021
\end{indented}

\begin{abstract}
The thermal conductivity of double-stranded (ds) B-DNA was systematically investigated using classical molecular dynamics (MD) simulations. The effect of changing base-pairs on the thermal conductivity of dsDNA, needed investigation at a molecular level. Hence, four sequences, viz. poly(A), poly(G), poly(CG) and poly(AT) were initially analysed in this work. Firstly, length of these sequences was varied from 4-40 base-pairs (bp) at 300 K and the respective thermal conductivity ($\mathrm{\kappa}$) was computed. Secondly, the temperature dependent thermal conductivities between 100 K and 400 K  were obtained in 50 K steps at 28 bp length. The M\"{u}ller-Plathe reverse non-equilibrium molecular dynamics (RNEMD) was employed to set a thermal gradient and obtain all thermal conductivities in this work. Moreover, mixed sequences using AT and CG sequencces, namely $\mathrm{A(CG)_{n}T}$ (n=3-7), $\mathrm{ACGC(AT)_{m}GCGT}$ (m=0-5) and $\mathrm{ACGC(AT)_{n}AGCGT}$ (n=1-4) were investigated based on the hypothesis that these sequences could be better thermoelectrics. 1-dimensional lattices are said to have diverging thermal conductivities at longer lengths, which violate Fourier law. These follow power law, where $\mathrm{\kappa \propto\  L^{\beta}}$.  At longer lengths, the exponent $\mathrm{\beta}$ need to satisfy the condition $\mathrm{\beta}>1/3$ for divergent thermal conductivity. We find no such significant Fourier law violation through divergence of thermal conductivities at 80 bp  lengths or 40 bp lengths. Also, in the case of second study, the presence of short (m$\le$2) encapsulated AT sequences within CG sequences show an increasing trend. These significant results are important for engineering DNA based thermal devices. 
\end{abstract}

\vspace{2pc}
\maketitle
\ioptwocol
\section{Introduction}

Theoretical investigations have shown DNA to have low thermal conductivity and might prove to be a better functional thermoelectric \cite{macia2007dna}. The earlier computational works have focussed on the thermal denaturation of B-DNA using Peyrard-Bishop-Dauxois (PBD) 1-dimensional (1D) model\cite{PeyrardPRL1989,Peyrardiop2004,VelizhaninPRE2011,Chieniop2013}. The heat conductivity of DNA double helix with only guanine sequence was computed using a coarse-grained model \cite{SavinPRB2011} as a deviation from these 1D models. Most models concentrate on the stable form of DNA, which is B-DNA\cite{Dickerson475}. A full-atomistic model of B-DNA with Drew-Dickerson sequence, ds(CGCGAATTCGCG) also showed heat capacity, thermal conductivity and phonon density of states of B-DNA  \cite{Mahalingam21}. A major insight, from experiments into the electrical and thermal conductivity of long $\lambda$-DNA fibers \cite{Kodama2009,Xuaip2014}, is obtained through the use of Scanning Tunnelling Microscopy (STM). Both quantum-tunneling and classical contribution of sequences become apparent in such experimental setups \cite{xu2004direct,li2016thermoelectric}.

Effective thermoelectrics are materials with high ZT, a dimensionless parameter expressed as

\begin{equation}
ZT = \frac{\sigma S^2T}{\kappa}\, \quad (\kappa = \kappa_{el} + \kappa_{l}),
\end{equation}
where at temperature T, electrical conductivity ($\sigma$), Seebeck coefficient (S) need to be high and thermal conductivity ($\kappa$) as a sum of electronic ($\kappa_{el}$) and lattice/phonon  ($\kappa_l$) conductivity need to be low. Materials with ZT$\approx$3 are better thermoelectrics than currently employed ones due to better conversion efficiencies \cite{ziabari2016nanoscale}. Whilst low dimensionality assists in achieving this, the search of materials has been restricted to Bismuth based alloys \cite{mao2020thermoelectric} or inorganic two-dimensional (2D) chalcogenides \cite{chen2017excellent,yu2016enhanced}. Organic thermoelectrics are seldom focussed. 

The thermal conductivity of B-DNA could be altered by engineering Adenine (A), Guanine (G), Cytosine (C) and Thymine(T) base-pair (bp) sequences. Hence, 4 sequences, viz. poly(A) or $(A)_n$, poly(G) or $(G)_n$, poly(CG) or $(CG)_m$ and poly(AT) or $(AT)_m$(n=4-40,m=n/2) were analysed for length dependence at 300 K. Additionally, at 28 bp length, the temperature dependence of $\kappa$ between 100 and 400 K in 50 K steps was investigated. A second set of computation were run on sequences  $A(CG)_{n}T$ (n=3-7), $ACGC(AT)_{m}GCGT$ (m=1-4) and $ACGC(AT)_{m-1}AGCGT$ (m=2-4) to observe the variance of thermal conductivities as a function of the sequence\cite{li2016thermoelectric}.

\section{Computational methods}

\subsection{Model setup}

\begin{figure}[h]
\centering
\includegraphics[width=1.25\linewidth]{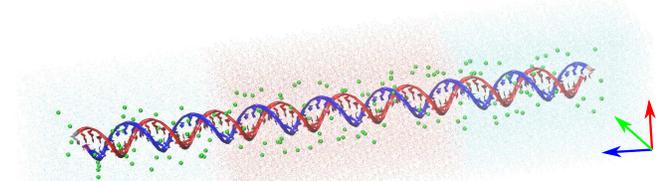}
\caption{Illustration showing a B-DNA sequence under thermal gradient with central hot (red) region and  cold (blue) regions at two ends. The green balls show Na$^+$ ions}
  \label{fgr:fig1}
\end{figure}

The B-DNA sequences (Figure \ref{fgr:fig1}) were generated using Nucleic Acid Builder (NAB) module in AMBERTOOLS18 \cite{caseamber18}. DNA OL15 force-field was used \cite{zgarbova2015refinement,galindo2016assessing}. A TIP3P water box was constructed using xleap module around the DNA such that a solvation shell of 15 \AA \ exists \cite{jorgensen1983comparison,price2004modified}. This is to prevent periodic image interference. Sodium (Na$^+$) ions were added to counter the negative phosphate(PO$_4^{3-}$) ions in the DNA backbone and make the system neutral \cite{joung2008determination}. The initial AMBER files were converted to LAMMPS data format and was used for simulation \cite{plimpton1995fast}. The minimization procedure begins by restraining the B-DNA with a force of 500 kcal/(mol\AA) . This minimizes the surrounding water. Then DNA was minimized by reducing the restraint from 20 kcal/(mol\AA) \ to 0 kcal/(mol\AA) \ in 5 cycles of the steepest descent and conjugate gradient minimization steps with step-size of 5000. Velocities were assigned to all atoms according to Maxwell-Boltzmann distribution for the required temperature. In order to make sure that DNA does not rotate or move while taking thermal conductivity calculation, the DNA is restrained throughout the simulation with a small force of 1 kcal/(mol\AA). SHAKE contraints were applied to hydrogen atoms as they decide the smallest timestep possible and might result in unstable simulation, if not applied \cite{ryckaert1977numerical,andersen1983rattle}. This enables a timestep of 1 fs. The system was equilibrated for 2 ns with Nos\'{e}-Hoover thermostat and 2 ns with Nos\'{e}-Hoover barostat with coupling constants 0.1 ps (100 timesteps) and 1.0 ps (1000 timesteps) \cite{shinoda2004rapid,tuckerman2006liouville}.  The long range interactions were taken care by particle-particle particle-mesh (PPPM) Ewald method\cite{hockney1988computer}

\subsection{Thermal conductivity calculation}

\begin{figure}[htbp]
\centering
  \includegraphics[width=\linewidth]{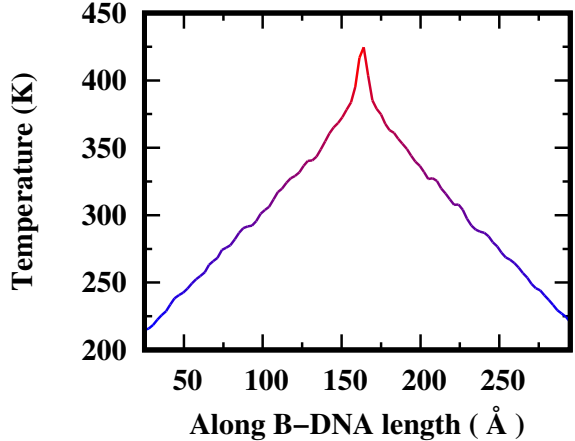}
  \caption{Temperature profile along length for a poly(A) sequence at 300 K}
  \label{fgr:fig2}
\end{figure}

M\"{u}ller-Plathe (MP) Reverse Non-Equilibrium method (RNEMD) was used for thermal conductivity calculation \cite{muller1997simple}. Here, a temperature gradient was set along the length of B-DNA and surrounding water box by swapping kinetic energies every step with the box being divided into 100 chunks in z-direction. With swapping kinetic energies every 20 fs, a temperature profile sets across DNA and surrounding water between the center hot region and cold regions on either side. The thermal conductivity $\kappa$ due to a linear temperature gradient between the DNA ends is

\begin{equation}
\kappa_{z}= \frac{\left( \frac{Q}{At} \right)}{\left(  \partial T / \partial z \right)},
\label{eqn:eqn2}
\end{equation}

where Q is the heat exchange between hot and cold regions, A is the cross-sectional area of the water box and t is the time for heat exchange. Also, note that the Q value above is already divided by 2 as heat flows from central hot chunk to 2 cold chunks in two directions.  First, an initial 1 ns basic run was used to setup a gradient. Then, a production run of 5 ns was used to generate statistical convergent data. The temperature gradient in equation~\ref{eqn:eqn2} is computed only across B-DNA from a smoothened temperature profile in this convergent regime. SHAKE restrained hydrogen atoms are excluded from calculations. A mild Berendsen thermostat with damping constant of 10000 timesteps is required to keep the system around desired temperature~\cite{zhang2005thermal}. 
\begin{figure}[!htbp]
\centering
  \includegraphics[width=\linewidth]{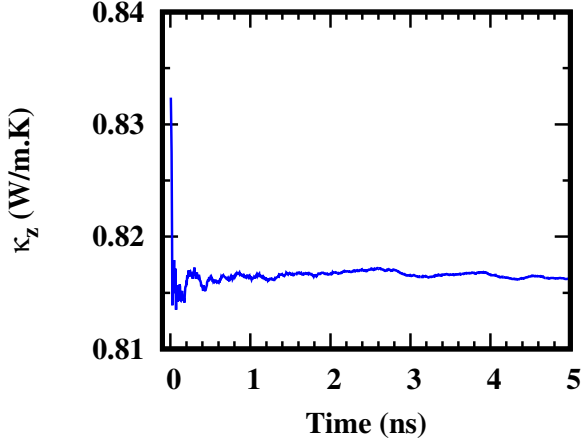}
  \caption{Sample convergence of thermal conductivity along length for a poly(A) sequence at 300 K}
  \label{fgr:fig3}
\end{figure}

The smoothened temperature gradient profile across a poly(A) B-DNA is shown in Fig. \ref{fgr:fig2}. After obtaining the temperature gradient, the  convergence of thermal conductivity with time is calculated (Fig. \ref{fgr:fig3}). All thermal conductivity data points plotted are obtained in the converged region as time averages of values between 4 ns and 5 ns.

\section{Results and discussion}

\begin{figure}[htbp]
\centering
  \includegraphics[width=\linewidth]{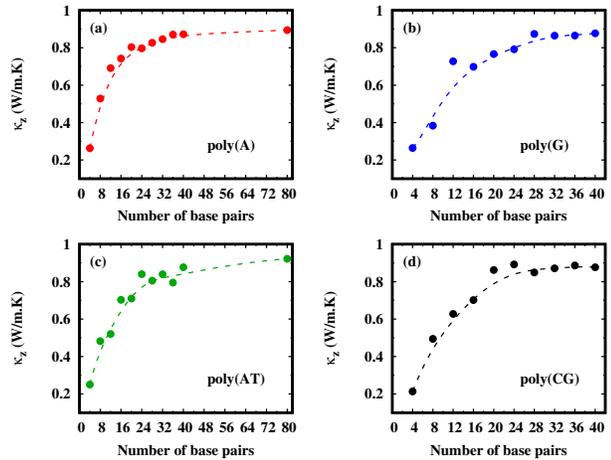}
  \caption{Length dependent thermal conductivity ($\mathrm{\kappa_z}$) along strand length  of (a) poly(A), (b) poly(G), (c) poly(AT) and (d) poly(CG) sequences at 300 K}
  \label{fgr:fig4}
\end{figure}

Firstly, the length dependence of the sequences poly(A), poly(G), poly(AT) and  poly(CG) at 300 K is analysed. The number of base pairs were increased in steps of 4 bp from 4 bp to 40 bp (Fig. \ref{fgr:fig4}). This resulted in the convergence of the thermal conductivities ($\mathrm{\kappa}$) of poly(G) and poly(CG) as the length reached 40bp. Whereas, poly(A) and poly(AT) sequences showed a linear increase from 32 to 40bp. Heat conduction in 1D chains and nanotubes have shown a power law ($\mathrm{\kappa\ \propto\  L^{\beta}}$) divergent dependence \cite{li2003anomalous,yang2010violation,xu2014length}, which violates Fourier law. In such anomalous situation, $\mathrm{\beta} \geq 1/3$. Hence, for these sequences $\mathrm{\kappa}$ at 80 bp was taken to resolve this question. It was found that $\mathrm{\beta}$ was close to zero for poly(A) and $\mathrm{\beta} \approx 0.13$ for poly(AT) sequence. This is nowhere close to the anomalous $\mathrm{\beta}$ and probably indicates that the convergence of the DNA thermal conductivity to a finite value at higher lengths. This finite $\mathrm{\kappa}$ value can be fitted  using the asymptotic relationship  

\begin{equation}
\mathrm{\kappa= \kappa_0 + \left(\frac{\kappa_1}{L}\right)}.
\label{eqn:eqn3}
\end{equation}
This accordance with Fourier law resulting in finite thermal conductivity has also been studied for 1D non-linear lattices \cite{SavinPRB2011,savin2014thermal,giardina2000finite}. The evidence here hints that heat transport in DNA follows normal diffusion in such repeating sequences.

Secondly, a moderately large DNA base-pair length was chosen (28 bp) and temperature dependence was calculated for the same sequences. One can observe that a decrease in thermal conductivity was observed long before the denaturation regime, where the sequence separates. No denaturation of the 4 sequences was observed as one increases from 150 K till 400 K.  All 4 sequences have their respective peak temperatures beyond which $\mathrm{\kappa}$ drops as a function of temperature. As in all solids, one can see that phonon-phonon scattering mechanism dominate and decrease the thermal conductivity in DNA sequence at temperatures higher than the peak temperature.
\begin{figure}[htbp]
\centering
  \includegraphics[width=\linewidth]{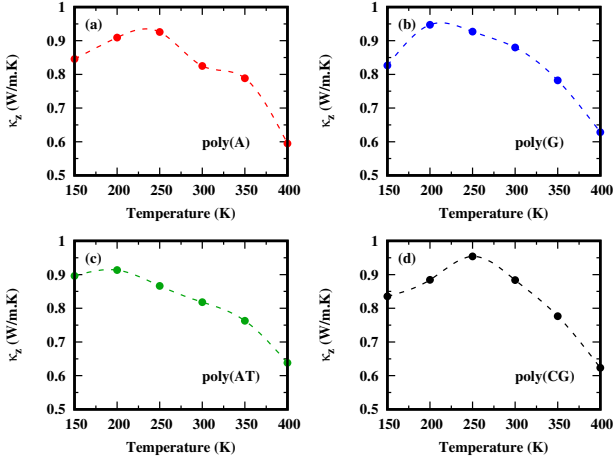}
  \caption{ Temperature dependent thermal conductivity ($\mathrm{\kappa_z}$) along strand length of (a) poly(A), (b) poly(G), (c) poly(AT) and (d) poly(CG) sequences at 28 bp length}
  \label{fgr:fig5}
\end{figure}

\begin{figure}[htbp]
\centering
  \includegraphics[width=\linewidth]{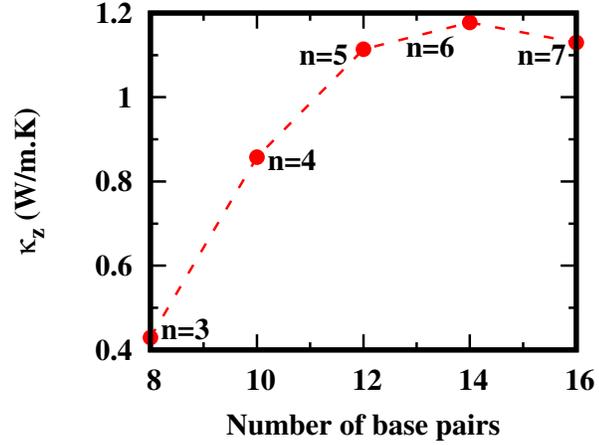}
  \caption{ Thermal conductivity ($\mathrm{\kappa_z}$) along strand length of sequences $\mathrm{A(CG)_{n}T}$ (n=3-7) }
  \label{fgr:fig6}
\end{figure}

\begin{figure}[htbp]
\centering
  \includegraphics[width=\linewidth]{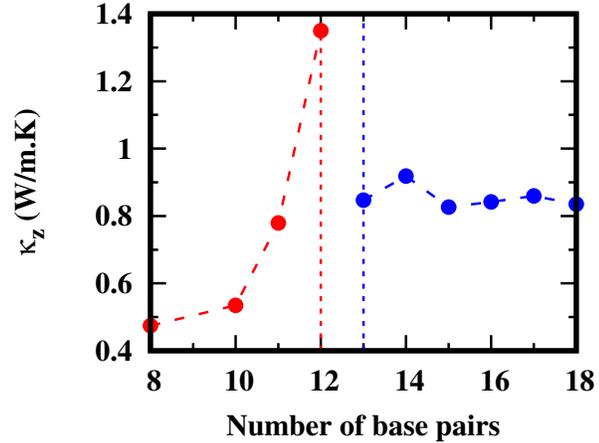}
  \caption{ Thermal conductivity ($\mathrm{\kappa_z}$) along strand length of sequences $\mathrm{ACGC(AT)_{m}GCGT}$ (m=0-5) and $\mathrm{ACGC(AT)_{n}AGCGT}$ (n=1-4) }
  \label{fgr:fig7}
\end{figure}

Sequences from a recent experimental work \cite{li2016thermoelectric} were chosen and their thermal conductivities were plotted at 300 K (Figures \ref{fgr:fig6} and \ref{fgr:fig7}). The $\mathrm{A(CG)_{n}T}$ sequences has a classical conduction mechanism similar to plain poly(CG) sequences (Figure \ref{fgr:fig4}). In contrast to ordinary poly(CG) sequences, encapsulating the poly(CG) sequences within Adenine (A) and Thymine (T) terminating sequences, increases the thermal conductivity by a slight margin.

The $\mathrm{ACGC(AT)_{m}GCGT}$ sequences showed an intriguing increase in thermal conductivity for m$\le$2 and n$\le$1 (red points in figure \ref{fgr:fig7}).  The mechanism changes and the thermal conductivity starts converging as length increases (blue points in figure \ref{fgr:fig7}). Fitting the red shows the possibility of an exponential dependence $\mathrm{\kappa\ \propto\  exp(\alpha \cdot L)}$ with $\mathrm{\alpha} \approx$ 1.053.  This is indeed interesting for a short change in length.

\section{Conclusion and outlook}

The thermal conductivity of B-DNA sequences found here hints at the convergences of DNA thermal conductivity as length increases for poly(A), poly(AT), poly(G) and poly(CG) sequence. Thermal switches using B-DNA is possible as the peak of $\kappa$'s temperature dependence is close to room temperature (Figure \ref{fgr:fig5}). The various sequences can be engineered to operate at different peak temperatures around room temperature. For example, poly(AG) sequences will have a peak between that of poly(A) and poly(G). Mixed (AT) and (CG) sequences have higher thermal conductivity than ordinary repeating sequences. The ratio of heat conduction through the phosphate backbone and base-pair might gives us more clues. The exponential dependence of (AT) blocks with (CG) blocks on either side, might be used to increase the thermoelectric figure of merit (ZT) of B-DNA. An experimental study on the thermal conductivity at smaller DNA length scales will further deepen the understanding in these directions.

\section*{Conflicts of interest}
There are no conflicts to declare.

\section*{Acknowledgements}
The authors acknowledge Ministry of Education, Government of India for financial support. 

\section*{References}
\bibliography{bdna}

\begin{thebibliography}{10}

\bibitem{macia2007dna}
Enrique Maci\'a.
\newblock Dna-based thermoelectric devices: A theoretical prospective.
\newblock {\em Phys. Rev. B}, 75:035130, Jan 2007.

\bibitem{PeyrardPRL1989}
M.~Peyrard and A.~R. Bishop.
\newblock Statistical mechanics of a nonlinear model for dna denaturation.
\newblock {\em Phys. Rev. Lett.}, 62:2755--2758, Jun 1989.

\bibitem{Peyrardiop2004}
Michel Peyrard.
\newblock Nonlinear dynamics and statistical physics of {DNA}.
\newblock {\em Nonlinearity}, 17(2):R1--R40, jan 2004.

\bibitem{VelizhaninPRE2011}
Kirill~A. Velizhanin, Chih-Chun Chien, Yonatan Dubi, and Michael Zwolak.
\newblock Driving denaturation: Nanoscale thermal transport as a probe of dna
  melting.
\newblock {\em Phys. Rev. E}, 83:050906, May 2011.

\bibitem{Chieniop2013}
Chih-Chun Chien, Kirill~A Velizhanin, Yonatan Dubi, and Michael Zwolak.
\newblock Tunable thermal switching via {DNA}-based nano-devices.
\newblock {\em Nanotechnology}, 24(9):095704, feb 2013.

\bibitem{SavinPRB2011}
Alexander~V. Savin, Mikhail~A. Mazo, Irina~P. Kikot, Leonid~I. Manevitch, and
  Alexey~V. Onufriev.
\newblock Heat conductivity of the dna double helix.
\newblock {\em Phys. Rev. B}, 83:245406, Jun 2011.

\bibitem{Dickerson475}
RE~Dickerson, HR~Drew, BN~Conner, RM~Wing, AV~Fratini, and ML~Kopka.
\newblock The anatomy of a-, b-, and z-dna.
\newblock {\em Science}, 216(4545):475--485, 1982.

\bibitem{Mahalingam21}
Vignesh Mahalingam and Dineshkumar Harursampath.
\newblock Thermal conductivity of b-dna.
\newblock {\em The Journal of Physical Chemistry B}, 125(5):1363--1368, 2021.

\bibitem{Kodama2009}
Takashi Kodama, Ankur Jain, and Kenneth~E. Goodson.
\newblock Heat conduction through a dna-gold composite.
\newblock {\em NANO LETTERS}, 9(5):2005--2009, MAY 2009.

\bibitem{Xuaip2014}
Zaoli Xu, Shen Xu, Xiaoduan Tang, and Xinwei Wang.
\newblock Energy transport in crystalline dna composites.
\newblock {\em Aip Advances}, 4(1):017131, 2014.

\bibitem{xu2004direct}
Bingqian Xu, Peiming Zhang, Xiulan Li, and Nongjian Tao.
\newblock Direct conductance measurement of single dna molecules in aqueous
  solution.
\newblock {\em Nano letters}, 4(6):1105--1108, 2004.

\bibitem{li2016thermoelectric}
Yueqi Li, Limin Xiang, Julio~L Palma, Yoshihiro Asai, and Nongjian Tao.
\newblock Thermoelectric effect and its dependence on molecular length and
  sequence in single dna molecules.
\newblock {\em Nature communications}, 7(1):1--8, 2016.

\bibitem{ziabari2016nanoscale}
Amirkoushyar Ziabari, Mona Zebarjadi, Daryoosh Vashaee, and Ali Shakouri.
\newblock Nanoscale solid-state cooling: A review.
\newblock {\em Reports on Progress in Physics}, 79(9):095901, 2016.

\bibitem{mao2020thermoelectric}
Jun Mao, Gang Chen, and Zhifeng Ren.
\newblock Thermoelectric cooling materials.
\newblock {\em Nature Materials}, pages 1--8, 2020.

\bibitem{chen2017excellent}
Kai-Xuan Chen, Shu-Shen Lyu, Xiao-Ming Wang, Yuan-Xiang Fu, Yi~Heng, and
  Dong-Chuan Mo.
\newblock Excellent thermoelectric performance predicted in two-dimensional
  buckled antimonene: a first-principles study.
\newblock {\em The Journal of Physical Chemistry C}, 121(24):13035--13042,
  2017.

\bibitem{yu2016enhanced}
Hulei Yu, Shuai Dai, and Yue Chen.
\newblock Enhanced power factor via the control of structural phase transition
  in snse.
\newblock {\em Scientific reports}, 6(1):1--12, 2016.

\bibitem{caseamber18}
DA~Case, IY~Ben-Shalom, SR~Brozell, DS~Cerutti, TE~Cheatham~III, VWD Cruzeiro,
  TA~Darden, RE~Duke, D~Ghoreishi, MK~Gilson, et~al.
\newblock Amber 2018; 2018.
\newblock {\em University of California, San Francisco}.

\bibitem{zgarbova2015refinement}
Marie Zgarbov{\'a}, Jiri Sponer, Michal Otyepka, Thomas~E Cheatham~III, Rodrigo
  Galindo-Murillo, and Petr Jure{\v c}ka.
\newblock Refinement of the sugar--phosphate backbone torsion beta for amber
  force fields improves the description of z-and b-dna.
\newblock {\em Journal of chemical theory and computation}, 11(12):5723--5736,
  2015.

\bibitem{galindo2016assessing}
Rodrigo Galindo-Murillo, James~C Robertson, Marie Zgarbova, Jiri Sponer, Michal
  Otyepka, Petr Jure{\v c}ka, and Thomas~E Cheatham~III.
\newblock Assessing the current state of amber force field modifications for
  dna.
\newblock {\em Journal of chemical theory and computation}, 12(8):4114--4127,
  2016.

\bibitem{jorgensen1983comparison}
William~L Jorgensen, Jayaraman Chandrasekhar, Jeffry~D Madura, Roger~W Impey,
  and Michael~L Klein.
\newblock Comparison of simple potential functions for simulating liquid water.
\newblock {\em The Journal of chemical physics}, 79(2):926--935, 1983.

\bibitem{price2004modified}
Daniel~J Price and Charles~L Brooks~III.
\newblock A modified tip3p water potential for simulation with ewald summation.
\newblock {\em The Journal of chemical physics}, 121(20):10096--10103, 2004.

\bibitem{joung2008determination}
In~Suk Joung and Thomas~E Cheatham~III.
\newblock Determination of alkali and halide monovalent ion parameters for use
  in explicitly solvated biomolecular simulations.
\newblock {\em The journal of physical chemistry B}, 112(30):9020--9041, 2008.

\bibitem{plimpton1995fast}
Steve Plimpton.
\newblock Fast parallel algorithms for short-range molecular dynamics.
\newblock {\em Journal of computational physics}, 117(1):1--19, 1995.

\bibitem{ryckaert1977numerical}
Jean-Paul Ryckaert, Giovanni Ciccotti, and Herman~JC Berendsen.
\newblock Numerical integration of the cartesian equations of motion of a
  system with constraints: molecular dynamics of n-alkanes.
\newblock {\em Journal of computational physics}, 23(3):327--341, 1977.

\bibitem{andersen1983rattle}
Hans~C Andersen.
\newblock Rattle: A “velocity” version of the shake algorithm for molecular
  dynamics calculations.
\newblock {\em Journal of computational Physics}, 52(1):24--34, 1983.

\bibitem{shinoda2004rapid}
Wataru Shinoda, Motoyuki Shiga, and Masuhiro Mikami.
\newblock Rapid estimation of elastic constants by molecular dynamics
  simulation under constant stress.
\newblock {\em Physical Review B}, 69(13):134103, 2004.

\bibitem{tuckerman2006liouville}
Mark~E Tuckerman, Jos{\'e} Alejandre, Roberto L{\'o}pez-Rend{\'o}n, Andrea~L
  Jochim, and Glenn~J Martyna.
\newblock A liouville-operator derived measure-preserving integrator for
  molecular dynamics simulations in the isothermal--isobaric ensemble.
\newblock {\em Journal of Physics A: Mathematical and General}, 39(19):5629,
  2006.

\bibitem{hockney1988computer}
Roger~W Hockney and James~W Eastwood.
\newblock {\em Computer simulation using particles}.
\newblock crc Press, 1988.

\bibitem{muller1997simple}
Florian M{\"u}ller-Plathe.
\newblock A simple nonequilibrium molecular dynamics method for calculating the
  thermal conductivity.
\newblock {\em The Journal of chemical physics}, 106(14):6082--6085, 1997.

\bibitem{zhang2005thermal}
Meimei Zhang, Enrico Lussetti, Lu{\'\i}s~ES de~Souza, and Florian
  M{\"u}ller-Plathe.
\newblock Thermal conductivities of molecular liquids by reverse nonequilibrium
  molecular dynamics.
\newblock {\em The Journal of Physical Chemistry B}, 109(31):15060--15067,
  2005.

\bibitem{li2003anomalous}
Baowen Li and Jiao Wang.
\newblock Anomalous heat conduction and anomalous diffusion in one-dimensional
  systems.
\newblock {\em Physical review letters}, 91(4):044301, 2003.

\bibitem{yang2010violation}
Nuo Yang, Gang Zhang, and Baowen Li.
\newblock Violation of fourier's law and anomalous heat diffusion in silicon
  nanowires.
\newblock {\em Nano Today}, 5(2):85--90, 2010.

\bibitem{xu2014length}
Xiangfan Xu, Luiz~FC Pereira, Yu~Wang, Jing Wu, Kaiwen Zhang, Xiangming Zhao,
  Sukang Bae, Cong~Tinh Bui, Rongguo Xie, John~TL Thong, et~al.
\newblock Length-dependent thermal conductivity in suspended single-layer
  graphene.
\newblock {\em Nature communications}, 5(1):1--6, 2014.

\bibitem{savin2014thermal}
Alexander~V Savin and Yuriy~A Kosevich.
\newblock Thermal conductivity of molecular chains with asymmetric potentials
  of pair interactions.
\newblock {\em Physical Review E}, 89(3):032102, 2014.

\bibitem{giardina2000finite}
Cristian Giardina, R~Livi, A~Politi, and M~Vassalli.
\newblock Finite thermal conductivity in 1d lattices.
\newblock {\em Physical review letters}, 84(10):2144, 2000.

\end{thebibliography}
\bibliographystyle{unsrt}
\end{document}